\documentclass[preprint]{elsarticle}

\usepackage{graphicx}% Include figure files
\usepackage{dcolumn}% Align table columns on decimal point
\usepackage{bm}% bold math
\usepackage[USenglish]{babel}
\usepackage[separate-uncertainty]{siunitx}% Nice micron signs and units
\usepackage{multirow}
\usepackage{textcomp}
\usepackage{hhline}
\usepackage{tabularx}
\usepackage[flushleft]{threeparttable}
\usepackage[table]{xcolor}
\usepackage{amssymb}
\usepackage{lipsum}
\usepackage{comment}
\usepackage[colorlinks=true, linkcolor=black, citecolor=prlblue, filecolor=black, urlcolor=prlblue]{hyperref}
\usepackage{caption}
\newcommand{\ee}{\hbox{$\rm{e^+}\rm{e^-}$}}

\def\ttdefault{cmtt}

\makeatletter
\def\bs{\expandafter\@gobble\string\\}
\def\lb{\expandafter\@gobble\string\{}
\def\rb{\expandafter\@gobble\string\}}
\def\@pdfauthor{C.V.Radhakrishnan}
\def\@pdftitle{elsarticle.cls -- A documentation}
\def\@pdfsubject{Document formatting with elsarticle.cls}
\def\@pdfkeywords{LaTeX, Elsevier Ltd, document class}

\DeclareRobustCommand{\LaTeX}{L\kern-.26em%
        {\sbox\z@ T%
         \vbox to\ht\z@{\hbox{\check@mathfonts
           \fontsize\sf@size\z@
           \math@fontsfalse\selectfont
          A\,}%
         \vss}%
        }%
     \kern-.15em%
    \TeX}
\makeatother

\journal{Physics Open}

\begin{document}

\begin{frontmatter}
\title{Proceedings of the Erice Workshop:\\ A new baseline for the hybrid, asymmetric, linear Higgs factory HALHF}

\author[1,2]{Brian~Foster\corref{cor1}}
\ead{brian.foster@physics.ox.ac.uk}
\cortext[cor1]{Corresponding author}
\author[3]{Erik Adli}
\author[4]{Timothy L. Barklow}
\author[2]{Mikael Berggren}
\author[5]{Stewart Boogert}
\author[3]{Jian Bin Ben Chen}
\author[1]{Richard~D'Arcy}
\author[3]{Pierre Drobniak}
\author[6]{Sinead Farrington}
\author[4]{Spencer Gessner}
\author[4]{Mark J. Hogan}
\author[3]{Daniel Kalvik}
\author[2]{Antoine Laudrain}
\author[3]{Carl~A.~Lindstr{\o}m}
\author[2]{Benno List}
\author[2]{Jenny List}
\author[7]{Xueying Lu}
\author[8]{Gudrid Moortgat Pick}
\author[2]{Kristjan P\~{o}der}
\author[9]{Andrei Seryi}
\author[3]{Kyrre Sjobak}
\author[2]{Maxence Th{\'e}venet}
\author[2]{Nicholas J. Walker}
\author[2]{Jonathan Wood}
\affiliation[1]{organization={John Adams Institute for Accelerator Science at University of Oxford},
addressline={Denys Wilkinson Building, Keble Road},
postcode={OX1 3RH},
city={Oxford},
country={United Kingdom}}
\affiliation[2]{organization={DESY},
addressline={Notkestrasse 85},
postcode={22607},
city={Hamburg},
country={Germany}}
\affiliation[3]{organization={Department of Physics, University of Oslo},
addressline={PO Box 1048, Blindern},
postcode={N-0316},
city={Oslo},
country={Norway}}
\affiliation[4] {organization={ SLAC National Accelerator Laboratory},
addressline={2575 Sand Hill Road},
postcode={CA 94025},
city={Menlo Park},
country={USA}}
\affiliation[5]{organization={Cockcroft Institute} ,
addressline={Daresbury Laboratory, STFC, Keckwick Lane, Daresbury},
postcode={WA4 4AD},
city={Warrington},
country={UK}}
\affiliation[6]{organization={Rutherford Appleton Laboratory} ,
addressline={STFC, Harwell Campus},
postcode={OX11 0QX},
city={Didcot},
country={UK}}
\affiliation[7]{organization={Argonne National Laboratory},
addressline={9700 S Cass Avenue},
postcode={IL60439},
city={Lemont},
country={USA}}
\affiliation[8]{organization={II. Institute of Theoretical Physics, University of Hamburg},
addressline={Luruper Chaussee 149},
postcode={22761},
city={Hamburg},
country={Germany}}
\affiliation[9]{organization={Thomas Jefferson National Accelerator Facility},
addressline={12000 Jefferson Avenue},
postcode={VA23606},
city={Newport News},
country={USA}}
\date{\today}
\begin{abstract}
The HALHF collaboration has discussed a new baseline for the project, taking into account comments from the accelerator community on various aspects of the original design. In particular, these concerned the practicality of the dual-purpose linac to accelerate both colliding positron bunches and the drive beams required for the plasma linac. In addition, many other aspects of the project were also considered; the discussion and conclusions are documented in this paper. Finally, a new baseline is outlined that has been optimised and addresses several weaknesses in the original design, has higher luminosity, reduced centre-of-mass energy boost and additional features such as positron polarization as well as electron polarization.  Although HALHF has become longer and more expensive, it remains  significantly smaller and cheaper than other mature Higgs factory designs currently under discussion.

\end{abstract}

\end{frontmatter}
%\maketitle
\section{Introduction}
\label{sec:intro}
\begin{figure*}[h!]
	\centering\includegraphics[angle=90, scale=1.4]{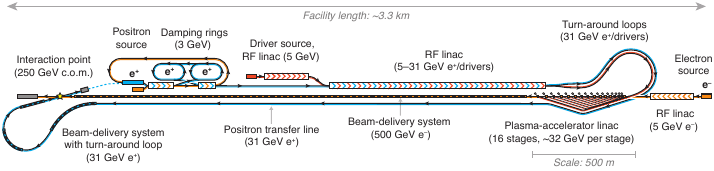}
	\caption{Schematic layout of the original baseline for the hybrid asymmetric linear Higgs factory HALHF. Particle sources provide electrons (orange), positrons (blue) and electron drivers (red) for acceleration. From Ref.~\cite{HALHF} (CC-BY 4.0).}
    \label{fig:1_setup}
\end{figure*}

The HALHF collaboration met for a workshop at the Ettore Majorana Centre, Erice, Sicily, in October 2024. Since the publication of the original HALHF concept~\cite{HALHF} and possible upgrades~\cite{HALHF_upgrades}, the collaboration has carried out significant work towards refining and optimising the design of the facility. The original layout is shown schematically in Fig.~\ref{fig:1_setup}. The purpose of the Erice 
workshop was to discuss salient points to establish a new baseline for the project, taking account of the work done and comments from the accelerator community. 

The collaboration has  not yet been able to consider all aspects of the HALHF facility in detail; indeed, some aspects, e.g.~the damping rings, have hardly been covered at all. Nevertheless, most aspects of the project have been refined since the original proposal. The discussions were initially held in plenary sessions, with overview talks to set the scene, followed by parallel working groups organised into the following broad topics:
\begin{enumerate}

    \item RF-linac designs: constraints and costs of drive-beam linac and positron linac;

    \item Beam-driven RF or structure wakefield acceleration (SWFA) as a cost-effective alternative for the positron linac;
    
    \item Beam-delivery system and positron source;
    
    \item The upgrade ladder: demonstrators, default energy, XCC~\cite{Barklow2022}, path to \SI{10}{TeV}
    
    \item Plasma-linac design I: staging, driver distribution;
    
    \item Plasma-linac design II: polarization, beam-quality preservation, tolerances; 
    
    \item Plasma generation, heating, cooling and power flow, efficiency; 
    
    \item Physics/detector design \& constraints, including coherent pairs. 
    
\end{enumerate}

This paper summarises in individual sections the discussions in each of the parallel-session working groups. The penultimate section reports on the result of these discussions in terms of the definition of a new baseline design for HALHF. The design is outlined and site layouts are illustrated. The final section provides a summary and outlook. 
\section{RF-linac designs: constraints and costs of drive-beam linac and positron linac}
\label{sec:linac}

\subsection{Separation of drive beam and positron linac}
\label{sec:separate}
One of the main weaknesses of the original HALHF proposal was the difficulty of designing the ``dual-purpose" linac to produce both electron drive beams of relaxed beam quality and simultaneously to accelerate a high-quality high-charge positron bunch to the required energy to produce \SI{250}{GeV} in the centre of mass (CoM). This stimulated examination of the possibility of separating these functions. 

Separating the positron linac from the drive-beam linac has a significant cost impact, as it adds a second linac to the design. However, it comes with a number of benefits:
first, it permits a separate optimisation of positron and drive-beam properties, in particular to construct a higher-gradient positron linac with increased positron beam energy, while operating the drive-beam linac at lower gradient. Also, the combined linac design accelerating two beams of opposite charge, different bunch charges and possibly different bunch lengths and emittances gives rise to complications for beam dynamics and beam steering as well as issues from beam loading that were not fully explored earlier; avoiding those complications is certainly beneficial to the design. However, the overall time structure (time between colliding bunches, number of bunches in the pulse, pulse repetition rate) has to be the same for drive beam and positron linac.

Separate linacs also allow removal of the positron turn-around, which is not scalable to high energy because of synchrotron-radiation losses and concomitant emittance growth, making emittance preservation easier. 
Overall, this design change was deemed beneficial and probably unavoidable, despite the cost increase it entails.

\subsection{Reduce the drive-beam bunch separation from 5~ns} 
The proposal to reduce the drive-bunch separation from \SI{5}{ns} to \SI{1}{ns} or less was motivated by the design of the chicanes in the plasma acceleration complex that make the drive-beam bunches  coincident with the witness bunch; the size of these chicanes scales with the bunch separation (see Sec.~\ref{sec:driver_dist}). This change proposal triggered a wider discussion in the linac working group.

In the simplest case, which was considered so far, the drive linac provides bunches at a constant rate $f = 1 / \Delta t_{\rm{db}}$, where  $\Delta t_{\rm{db}}$ is the bunch spacing, forming a pulse of length $t_p = n_b n_s \Delta t_{\rm{db}}$, with $n_b$ and $n_s$ being the number of colliding bunches in a pulse and the number of stages, respectively.
While short pulse lengths are beneficial from a linac-efficiency viewpoint, as they reduce the time when RF power heats the cavity wall and is thus lost for beam acceleration, the beam current and thus the instantaneous power rise as $f$.
Operation in a mode where klystrons provide the full instantaneous beam power of the drive-beam pulse then leads to an extraordinary large amount of klystrons that dominate the accelerator cost for a \SI{\sim1}{ns} bunch spacing.
In a configuration where klystron cost dominates, a plasma stage, which transfers around 40\% of the drive beam power to the witness beam, will not be more cost effective than an accelerator that uses klystrons to accelerate a witness beam directly.

It is necessary to reduce the peak power of the klystrons to a manageable and affordable value. This requires a lengthening of the RF pulse length, and some sort of manipulation that concentrates the energy provided by the RF pulse into the pulse of 
drive-beam bunches that accelerate one witness bunch. Possible means to achieve this are RF pulse compression, manipulation of the drive-beam bunch pattern with delay loops and combiner rings, or the use of the RF cavities themselves as an energy buffer, which would favour a superconducting RF linac.

Generically, the working group concluded that the time between colliding bunches $\Delta t_{\text coll}$ should be decoupled from $n_s\Delta t_{db}$, which introduces a new degree of freedom into the optimisation. 
Two ideas were presented:
\begin{enumerate}
    \item	For a normal-conducting linac, in particular one based on travelling waves,  use delay loops and combiner rings similar to those of the Compact Linear Collider (CLIC) proposal~\cite{Charles2018} to reduce the  spacing of drive-beam bunches from e.g.~\SI{4}{ns} in the drive linac to \SI{2}{ns} or smaller at the end of the combiner rings. However, the circumference of such rings would favour a significant reduction in drive-beam energy;
    \item Use superconducting L-band cavities at an International Linear Collider (ILC)-like frequency such as \SI{1.3}{GHz}. An L-band cavity at \SI{20}{MV/m} stores sufficient energy to accelerate a group of 16 bunches  with tolerable voltage drop~\cite{NW_Oslo}.
\end{enumerate}

\subsection{Reduced energy asymmetry}
\label{sec:easym}
The separation of drive beam and positron linac opens up a new design choice to optimise drive beam and positron energies separately. A higher positron energy than \SI{31}{GeV} seems to be preferred, based on a cost--benefit analysis of providing more energy via klystrons to the colliding positron beam compared to more to the drive beams and hence to the plasma accelerator for the colliding electron beam.  

\subsection{Larger number of stages in the plasma linac}
A larger number of stages leads to more bunches in the drive-beam linac, drawing out the time over which RF power is consumed. If this leads to a reduction of peak power consumption and thus the required number of klystrons, it is a beneficial change.

\subsection{Higher transformer ratio in the plasma linac}
A higher transformer ratio in the plasma cells means that drive-beam energy and gradient can be reduced in favour of increased drive-beam bunch charge. However, large bunch charges produce significant challenges for the drive-beam linac because of the energy spread caused by longitudinal wakefields. Those longitudinal wakefields are counteracted in the linac by the longitudinal voltage variation from running the cavities off-crest, an effect that is reduced if the gradient and/or the frequency are reduced. Increasing the transformer ratio to 2 while running at a higher gradient and with a somewhat smaller bunch charge than that of the CLIC drive-beam linac is both practical and beneficial.
 
\subsection{Longer bunch trains}
Increasing the number of colliding bunches within one pulse implies that the total beam energy delivered in one pulse is increased, directly increasing the delivered luminosity but leading to a corresponding increase in modulator cost and volume. Depending on other beam parameters, dynamic heating of the cavities and/or the klystrons may be a factor limiting the pulse length.

\subsection{Undulating delay chicanes}
As discussed above, the design of delay chicanes determines the required separation of drive-beam bunches. Going significantly below \SI{5}{ns} drive-beam bunch separation would pose challenges if kickers were used in the drive-beam linac but this can be avoided by instead utilising RF deflectors.

\subsection{Drive-beam linac parameters}
A normal-conducting, room-temperature L-band (\SI{1}{GHz}) travelling-wave linac with low (2--\SI{4}{MV/m}) gradient and \SI{<5}{GeV} beam energy, similar to the CLIC drive-beam design, seems optimal. This appears possible; however, component costs for L-band equipment (cavities and klystrons) are expected to be higher than for S-band (\SI{3}{GHz}) components owing to the overall larger dimensions. 
Preliminary estimates indicated that \SI{1250}{\micro\meter}-long triangular bunches with an \SI{8}{nC} bunch charge at \SI{4}{MV/m} and \SI{5}{\degree} RF phase would result in a 1.3\% energy spread. 

An S-band design might be feasible for the intended bunch charges and bunch lengths, assuming that requirements on the drive-beam energy spread are lower than for colliding bunches.
Overall, a better understanding of the requirements from the plasma-acceleration stages on the drive-beam bunches in terms of transverse emittance, bunch length, longitudinal charge profile and energy spread is needed for realistic simulations of the drive-beam linac.

\subsection{Positron-linac parameters}
A high-gradient e.g.~40--\SI{60}{MV/m} S-band (\SI{3}{GHz}) travelling-wave linac, possibly running at cryogenic temperatures similar to that proposed for the $\rm{C}^3$ project~\cite{C3}, would be suitable to produce the high-energy high-quality positron beam required for HALHF. Owing to the large gradient that counteracts longitudinal wake fields, this appears feasible even for the high bunch charge proposed.

\section{Beam-driven RF/SWFA as a cost-effective alternative for the positron linac}
\begin{figure*}[t]
	\centering\includegraphics[width=\textwidth]{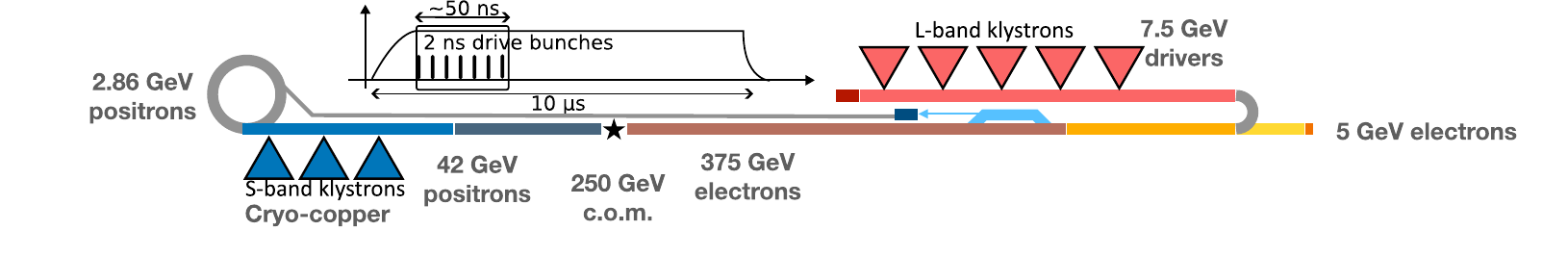}
	\caption{Schematic concept and drive-beam timing for a typical scheme for HALHF. Drive beams from an L-band linac (shown in pink) drive the PWFA linac (orange) which passes through a wiggler (light blue) to generate positrons before travelling to the interaction point (IP) (star) via the beam-delivery system (BDS) (brown). Positrons are transported to damping rings (gray) and then accelerated in an S-band linac (blue), e.g.~Cooled Copper, before passing through a BDS (dark blue) before colliding at the IP.}
    \label{fig:2_positron_time}
\end{figure*}

HALHF uses RF structures for positron acceleration because of the difficulty in applying beam-driven plasma-wakefield acceleration (PWFA) technology to positively charged particles~\cite{Cao2024}. Using a separate linac and increasing the positron energy makes high gradient desirable. Furthermore, the time structure of the positron beam must be the same as the PWFA electron beam, as illustrated in Fig.~\ref{fig:2_positron_time} for a typical option with separated linacs. The total pulse length must therefore be relatively long, on the order of \SI{10}{\micro\s} (assuming no combiner rings). The beam to be accelerated is composed of high-charge, low-emittance bunches being accelerated at intervals of \SI{\approx 50}{ns}. This is a challenging parameter set for an RF accelerator, and also requires a large number of klystron--modulator units to drive it.

\begin{figure*}[t]
	\centering\includegraphics[width=\textwidth]{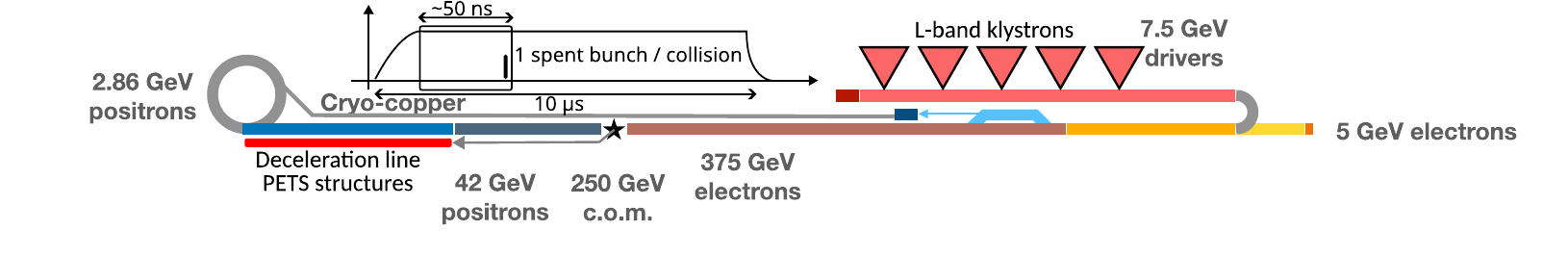}
	\caption{Schematic concept and HALHF machine layout and timing for SWFA with spent colliding beam reuse. The spent electron beam is focused and guided to drive power extraction and transfer structures (PETS) (shown in red) that provide power for the positron linac. Other details as in the caption to Fig.~\ref{fig:2_positron_time}.}
    \label{fig:3_spent}
\end{figure*}

\begin{figure*}[t]
	\centering\includegraphics[width=\textwidth]{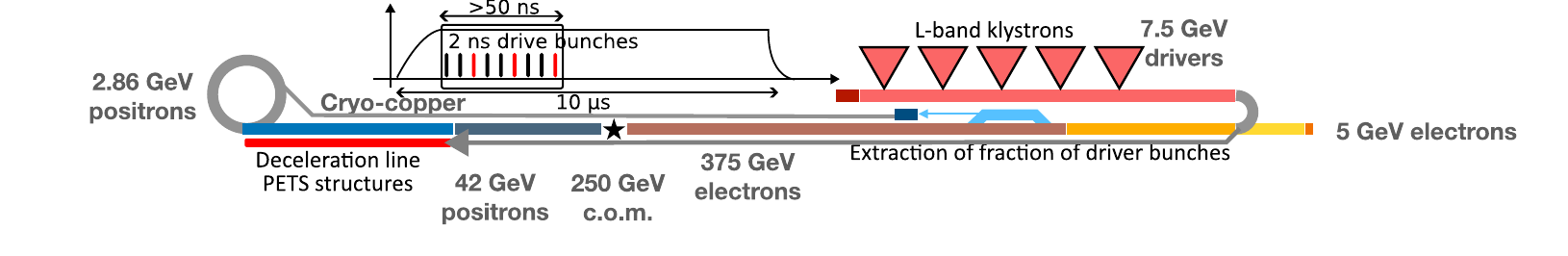}
	\caption{Schematic concept and HALHF machine layout for one example of a ``continuous'' drive-beam timing scheme using a beam-transformer SWFA positron accelerator scheme, with 2:1 interleaved multiplexing PWFA:SWFA. A fraction of the drive bunches are extracted before entering the PWFA accelerator and used to drive power extraction and transfer structures (PETS) that provide power for the positron linac. Other details as in the caption to Fig.~\ref{fig:2_positron_time}.}
    \label{fig:4_two_structure}
\end{figure*}

\begin{figure*}[t]
	\centering\includegraphics[width=\textwidth]{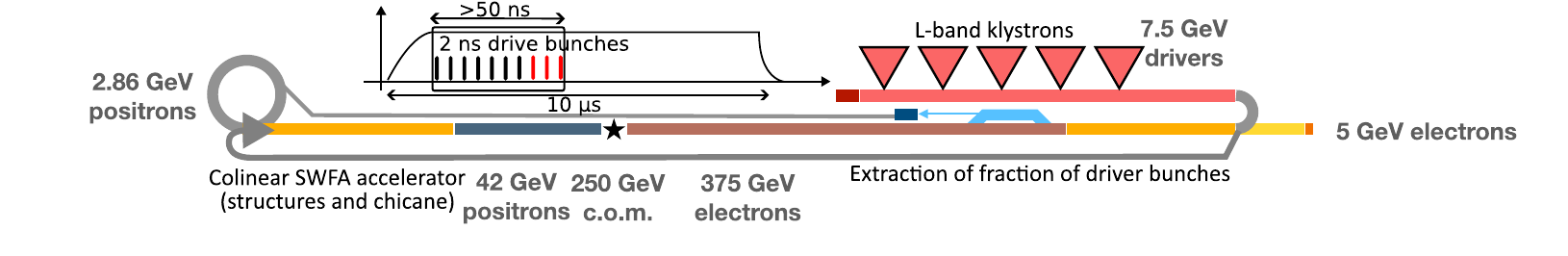}
	\caption{Schematic concept and HALHF machine layout for one example of a ``burst'' drive beam timing scheme with 7:3 multiplexing in a collinear SWFA positron accelerator scheme, which is very similar to the burst beam transformer scheme. One extra drive-beam turn-around is needed. Other details as in the caption to Fig.~\ref{fig:2_positron_time}.}
    \label{fig:5_collinear}
\end{figure*}

Using SWFA to avoid the expense and upkeep of klystrons and modulators by using an already existing beam to power the positron accelerator is an attractive alternative option. This also opens up the possibility of RF pulse shapes different to those that are feasible with conventional power sources, which could enhance the accelerating gradient. Several different options have been considered for this: the reuse of the spent electron beam (see Fig.~\ref{fig:3_spent}); a two-structure beam transformer similar to CLIC~\cite{CLIC_CDR_2012, CLIC_PIP_2018} (see Fig.~\ref{fig:4_two_structure}) and a collinear SWFA accelerator~\cite{Zholents2018,Jing2022} (see Fig.~\ref{fig:5_collinear}).

For a two-beam beam-transformer scheme, the drive beam is decelerated in a power extraction and transfer structure (PETS) to provide RF power that is used as a power source for the RF accelerating structure for the positron beam. This concept was investigated by the CLIC project at the CTF-3 facility~\cite{Ruber2013}, demonstrating acceleration of a probe beam in a normal-conducting X-band structure at gradients in excess of \SI{100}{MV/m}, and production of RF power above \SI{100}{MW}. Recent work at the Argonne Wakefield Accelerator (AWA) has demonstrated over \SI{500}{MW} of power extraction~\cite{Picard2022} and gradients about \SI{300}{MV/m}~\cite{Shao2022}, with however much shorter RF pulses.

The amount of power produced in a PETS structure at steady state is given by~\cite{Adli2009,Adli2011}
\begin{equation}
    P=\frac{R'}{4Q}\frac{\omega_{RF}}{v_g }(LIF\eta_\Omega)^2,
\end{equation}
where $R'/Q$ is the PETS linac impedance per unit length, $\omega_{RF}$ is the angular frequency of the excited mode in the structure, $v_g$ the group velocity, $L$ the structure length, $I$ the average beam current, $F$ the form factor determined by the bunch longitudinal profile, and $\eta_\Omega$ the ohmic loss efficiency. However, to reach this steady state, the PETS must be fed a constant stream of bunches during the RF output filling time $t_\mathrm{fill}=L(1-v_g/c)/v_g$; it is therefore not effective to drive the PETS with a single bunch since the field from the previous bunches is required to extract energy efficiently from the subsequent bunches, unlike the situation in PWFA, where the bunch length is long relative to the wavelength.

The requirement of drive-bunch trains for efficient two-beam acceleration excludes the reuse of the spent electron beam from the post collision line (see Fig.~\ref{fig:3_spent}) since the average current would be low, and a long filling time would be required – requiring very long power extraction and transfer structures (PETS). Another challenge is the need to produce multiple high-energy electron beams that would pass through the interaction point (IP) without colliding, in order to power the positron structures before a single positron bunch can be accelerated. Otherwise this would be a tempting option because it reuses kinetic energy that would otherwise be converted to heat at the beam dump.

In a two-structure beam-transformer scheme similar to CLIC, the drive beam would be the same as for the PWFA, as illustrated in Fig.~\ref{fig:4_two_structure}, but with a different average current (i.e., a different bunch spacing). The drive beam can be time-multiplexed in two ways: either with interleaved bunches at some factor corresponding to the desired current and bunch frequency in the PETS as in Fig.~\ref{fig:4_two_structure}, or with a part of the first or last half of the driver bunch train sent to the positron linac PETS as in the collinear scheme shown in Fig.~\ref{fig:5_collinear}. The second option might require an extra turn-around. For the first variant, the same type of RF structures as in the baseline could potentially be used, with RF power provided by PETS instead of klystrons. For the second variant using the timing scheme illustrated in Fig.~\ref{fig:5_collinear}, the RF power would be modulated on/off once per bunch crossing. A potentially attractive feature of this for high gradient is that the RF structures do not stay powered longer than necessary, which could enable higher accelerating gradient if the wakefields associated by higher R/Q and frequency structures, which have smaller apertures, can be tolerated. A challenge for the second scheme is the efficiency, as both the PETS and main positron-beam structures must be filled before a positron bunch could be accelerated.

A collinear structure wakefield accelerator, as illustrated in Fig~\ref{fig:5_collinear}, is similar to the ``burst'' beam transformer scheme discussed above, except both beams travel in the same structure. This reduces the number of components, but requires two beams with different energies and stability requirements to be transported in the same beamline and RF cavities, as is the case for PWFA. Furthermore, the fundamental theorem of beam loading~\cite{Wilson1977} applies, requiring a  driver bunch shaped to increase the transformer ratio of the accelerating:decelerating voltages above 2:1 to improve acceleration efficiency~\cite{Zholents2018}.

In summary, an SWFA scheme in which the positron linac is also driven by the drive beam linac could work well for HALHF, since the drive-beam linac is required anyway. It would however increase the complexity of the machine as well as introducing more non-conventional technologies, potentially increasing the risk. The use of SWFA is therefore an option after future study and optimisation rather than a choice for a baseline.

\section{Beam-delivery system and positron source}
\label{sec:bds-positron}

\subsection{Beam-delivery system}
The design of the HALHF beam-delivery system (BDS) strongly benefits from Next Linear Collider (NLC) and ILC BDS designs~\cite{Seryi2004, Seryi2007}, as well as that for CLIC. The BDS for HALHF will need all of the sub-systems included in the NLC/ILC design, \textit{viz.}~a coupling correction and diagnostics section, polarimetry and emergency extraction beamline, betatron and energy collimation, an energy spectrometer and final-focus system. However, further optimisation of the design, specifically for HALHF parameters and approaches, is possible.

First, for HALHF with a normal-conducting linac and corresponding short pulse train, the existing design closest to HALHF parameters is the NLC BDS with renewable collimation spoilers~\cite{Markiewicz2014} (in contrast to survivable spoilers in the ILC BDS). The engineering design of renewable spoilers will need to be further advanced during the technical design phase of HALHF. 
From the point of view of beam tests of the final focus, ATF2 at the KEK Final Focus~\cite{Seryi2014, White2014} test beam is both an essential and sufficient demonstration. 
    
While the NLC BDS design is a good start for the HALHF BDS, optional design improvements for HALHF are possible. Specifically, distributed collimation inside and between plasma sections should be considered to clean-up beam tails as near as possible to where they are created (see Sec.~\ref{sec:collimation}). This distributed collimation will also likely ease the design of renewable BDS collimators. Integration of distributed collimation requires coordination between the BDS and plasma-acceleration teams of the HALHF collaboration. Optimisation of the BDS design, and specifically its collimation system and interaction-region magnets, will also need to take into account the asymmetry of electron- and positron-beam emittances and bunch charges. BDS optimisation will feed back into the overall HALHF design.  
    
One of the most attractive features of HALHF is its staged upgradability to TeV and possibly multi-TeV energies~\cite{HALHF_upgrades}. Discussion of HALHF upgrades to TeV and multi-TeV opened up a plethora of ideas. While the final focus itself scales well to multi-TeV energies~\cite{Raimondi2001}, the most unfavourable scaling is for energy collimation. However, novel ideas such as distributed collimation in plasma sections and nonlinear energy collimation~\cite{RestaLopez2011} should be studied, as they will be likely to restore favourable scaling and allow a compact BDS for multi-TeV advanced colliders to be designed. A very important parameter for the HALHF design, which may either ease or complicate a TeV and multi-TeV upgrade, is the collision crossing angle, and specifically the angle between linacs, which cannot be changed once the tunnels are constructed. While  NLC, CLIC and eventually ILC designs had large crossing angles of 14--\SI{20}{mrad}, the recently suggested FEL-based $\gamma\gamma$ multi-TeV collider concept XCC~\cite{Barklow2022} requires a small crossing angle of just a couple of mrads. Maintaining flexibility of BDS design for Higgs energies while also enabling such multi-TeV upgrades might require re-consideration of the crossing-angle assumption and an analysis of the applicability of earlier designs of zero~\cite{TESLA-TDR-II} or 2-mrad~\cite{Moffeit2006}~crossing-angle BDS systems, or a flexible dual BDS~\cite{Cilento2021}.

\subsection{Positron source}

\begin{figure}[ht]
\centering
\includegraphics[width=\linewidth]{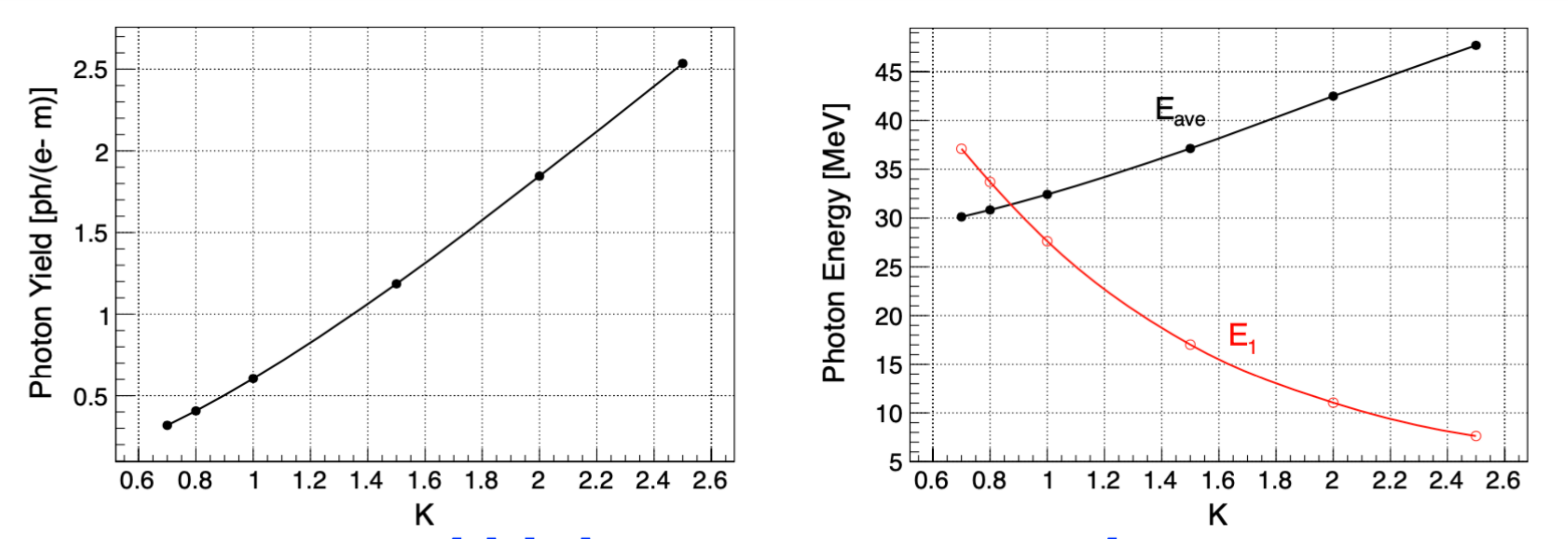}
\caption{Photon yield (left) and photon energy (right) as a function of
the undulator $\kappa$-value. $E_{\mathrm{ave}}$ is the average photon energy and $E_1$ 
is the energy cut-off of the 1st harmonic. From Ref.~\cite{Ushakov:2013bm}.}
\label{fig:6_undulator}       
\end{figure}

\begin{figure}
\centering
\includegraphics[width=\linewidth]{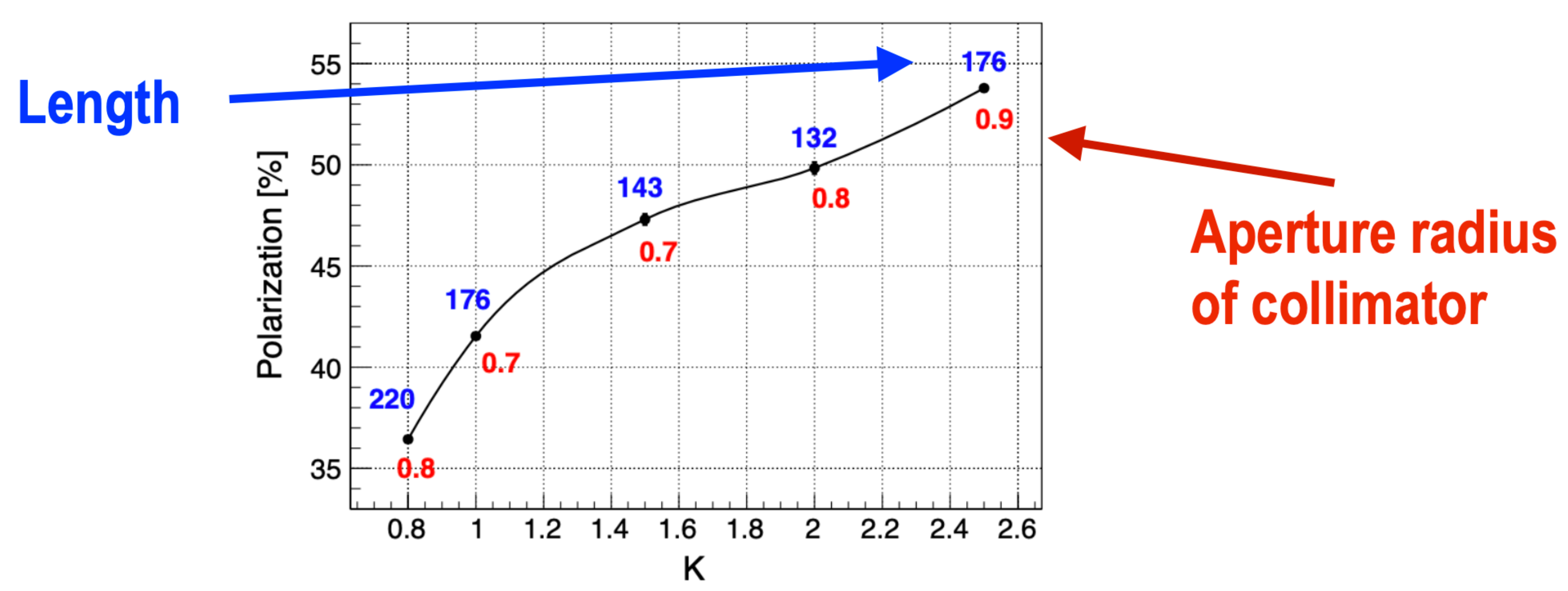}
\caption{Positron polarization versus $\kappa$-value without collimation but with an optic-matching device generating a peak field on the beam axis of \SI{3.2}{T}. The blue numbers indicate the required undulator length. From Ref.~\cite{Ushakov:2013bm}.}
\label{fig:7_polarization}      
\end{figure}

The positron source for HALHF, and indeed for any \ee\ linear collider, is challenging. This is particularly the case when, for very strong  reasons related to physics reach~\cite{Moortgat-Pick:2015lbx}, polarized positrons are essential. Fortunately, a source with similar characteristics to that required for HALHF has already been substantially designed for ILC. 

In the ILC design, a \SI{125}{GeV} ${\rm{e}^-}$-beam passes through a superconducting helical undulator, generating circularly polarized photons with energy \SI{\sim7.5}{MeV} that impinge on a thin rapidly rotating target constructed from a Ti alloy, producing polarized electrons and positrons. The positrons are captured, pre-accelerated and led through spin rotators before entering damping rings. Advanced simulations of the undulator have been performed~\cite{Alharbi:2024dud}, e.g.~examining the impact of field misalignments, errors in the magnetic field ($\kappa$-value) and the period $\lambda$. Depending on the $\kappa$-value, such uncertainties can affect the attainable polarization and the  load on the target.

The following subsections describe work carried out to strengthen the ILC design and then its adaption for HALHF. 

\subsubsection{Rotating target}
The \SI{1}{m}-diameter target for ILC was foreseen to rotate on magnetic bearings at \SI{2000}{rpm}, which corresponds to a tangential speed of \SI{100}{m/s}. This leads to the photon beam returning to the same target position every 7 seconds. The beam has a power of \SI{\sim60}{kW}, but only about 3\% of this is deposited in the target. Radiative cooling is sufficient within the vacuum chamber~\cite{Riemann:2020ytg}; the heat from the vacuum chamber is taken away by water cooling. Discussions on manufacturing the device are ongoing with the SKF company in Canada. 
 
\subsubsection{Mask design} 
A mask system to protect the undulator walls has been studied, designed to restrict  the synchrotron radiation deposition to \SI{<1}{W/m}, even at the maximal length of \SI{320}{m} for the ILC. 

\subsubsection{Optics-matching device}
There has recently been substantial progress towards manufacturing a scaled-down prototype pulsed solenoid used as an optics-matching device, suitable for \SI{1}{ms} photon pulses hitting the target. Manufacturing drawings have been produced and prototypes manufactured using 3D-printing. First measurements of the fields with \SI{1}{kA} (pulsed and DC) are planned in 2025, which will be extended to \SI{50}{kA} at CERN. The higher yield required for HALHF (estimated to 3--4 positrons/electron), greater than the 1.5~positrons/electron required for ILC, should be achievable and is currently being studied~\cite{Moortgat-Pick:2024fcy}. 

\subsubsection{Undulator-based positron source for HALHF}
\label{sec:undulator-HALHF}
The fully accelerated HALHF $e^-$ beam had an energy of \SI{500}{GeV} in the original baseline. This would be used as the photon source in the undulator; this has much higher energy than the \SI{125}{GeV} used in the ILC Higgs-factory design, so that the undulator parameters require adjustment.

The parameters calculated for the original \SI{1}{TeV} ILC upgrade-option~\cite{Ushakov:2013bm} can be used as a starting point, i.e.~an undulator with \SI{174}{m} length, a period of $\lambda=\SI{43}{mm}$ and a high magnetic field with $\kappa=2.5$.
The undulator radiation is simulated using Kincaid's formulae~\cite{Kincaid:1977fg}.
The photon-generation efficiency in such an undulator as a function of 
$\kappa$ is shown in the left-hand plot of Fig.~\ref{fig:6_undulator}, where the photon yield has been normalised per electron per meter of undulator. The photon energy cut-off of the 1st harmonic and the average photon energy are shown in the right-hand plot. The impact of the undulator field on the $\rm{e}^+$ polarization is shown in Fig.~\ref{fig:7_polarization}.

A parameter set such as that for ILC at \SI{1}{TeV} could produce polarization of up to 54\%.  However, the high $\kappa$ value means that higher harmonics are important, leading to higher mean power and greater energy spread for the photon beam. This makes the $e^+$ capture more difficult but the use of e.g.~a pulsed solenoid or indeed a plasma lens give grounds for optimism. 

Detailed simulations using \textit{CAIN} adapted for undulator radiation~\cite{Yokoya} are currently being carried out using the above scheme for HALHF. It is expected that estimates for the achievable HALHF positron yield will soon be available.

\section{The upgrade ladder: demonstrators, default energy, XCC, path to 10 TeV}

\subsection{Mapping of required R\&D items for electroweak-scale HALHF upgrades onto existing beam test facilities and future demonstrators}
HALHF is designed for an initial center-of-mass energy of \SI{250}{GeV}, but the full physics programme at a linear-collider facility should include studies of the $t \overline{t}$ threshold at \SI{380}{GeV} and the Higgs self-coupling at \SI{550}{GeV}. HALHF can be extended to reach these higher energy targets without drastic changes to the facility concept. Table~\ref{tab:1} shows the beam energies assuming a constant boost of $\gamma = 2.13$ as in the original baseline~\cite{HALHF_upgrades}: 

\begin{figure}[ht]
    \centering
    \includegraphics[width=\linewidth]{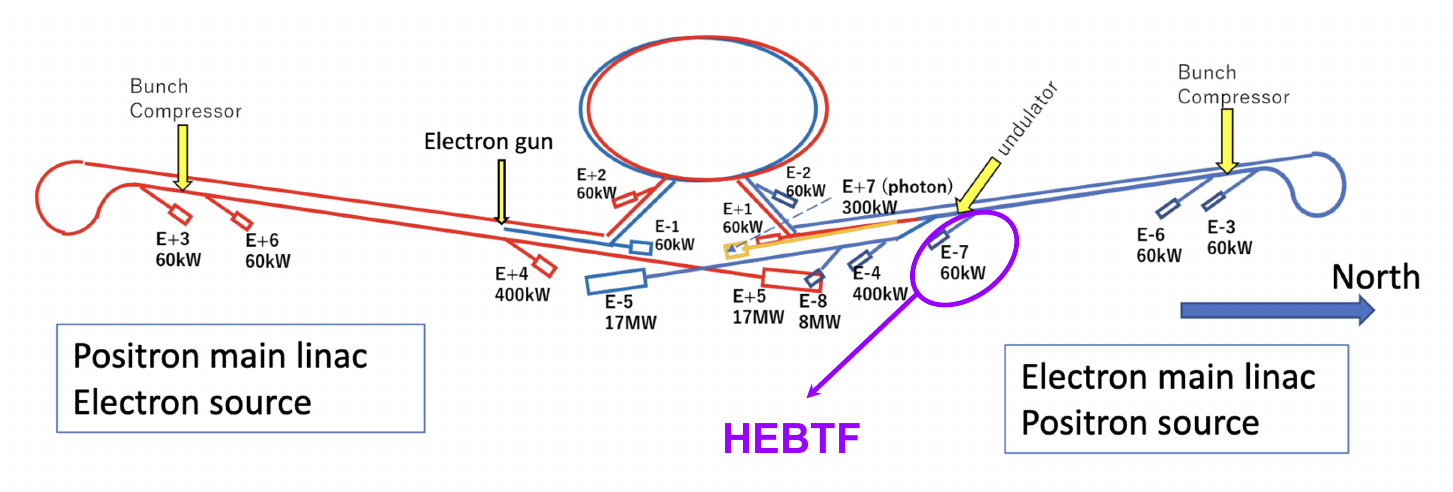}
    \caption{The figure illustrates a possible location for a High Energy Beam Test Facility (HEBTF) at the ILC (adapted from Fixed Target Experiments at the ILC, ILCX Workshop 2021). The red and blue lines illustrate the paths of positron and electron bunches, respectively, to beam dumps labelled ``E+n" for the $n$th dump. Instead of being dumped in E-7, the beam would enter the HEBTF beamline.}
    \label{fig:8_HEBTF}       
\end{figure}

\begin{table}[!hbtp]
    \begin{tabular}{p{0.25\linewidth}>{\centering}p{0.19\linewidth}>{\centering}p{0.19\linewidth}>{\centering\arraybackslash}p{0.17\linewidth}}
    Center-of-mass energy & $e^-$ beam energy & $e^+$ beam energy& Facility length \\[15pt]
    \hline \\[-7pt]
     250 GeV & 500 GeV & 31 GeV & 3.3 km \\
     380 GeV & 760 GeV & 47.5 GeV & 3.7 km \\
     550 GeV & 1.1 TeV & 68.5 GeV & ~5 km \\[3pt]
    \hline
    \end{tabular}
   \caption{Original HALHF baseline parameters at different center-of-mass energies with constant $\gamma = 2.13$, as estimated in Ref.~\cite{HALHF_upgrades}.}
   \label{tab:1}
\end{table}

Both the \SI{250}{GeV} centre-of-mass (CoM) HALHF concept and the EW-scale upgrades require a staging demonstration facility. The SPARTA staging demonstrator is envisaged as a compact machine based on plasma acceleration that uses few-GeV drivers to accelerate a main beam up to tens of GeV energy~\cite{Lindstrom2022}. The 20+ GeV energy scale is relevant for strong-field quantum electrodynamics (SFQED) studies and is greater than the electron-beam energy available at FACET-II for E320~\cite{Reis2024} or XFEL for LUXE~\cite{Abramowicz2024}.

A SPARTA PWFA staging demonstrator with an electron--laser interaction region for SFQED studies also provides an R\&D platform for a future $\gamma\gamma$ collider such as XCC~\cite{Barklow2022}. In fact, a SPARTA demonstrator that can reach 60+ GeV beam energy would be equivalent to one arm of a PWFA-based XCC concept. A successful R\&D programme for SPARTA/XCC would include a demonstration of electron/X-ray scattering and operation at high repetition rate. Such a facility could naturally evolve into a \SI{125}{GeV} CoM $\gamma\gamma$ collider and become the first stage of a HALHF Higgs Factory.

\subsection{Mapping of required R\&D items for multi-TeV-scale HALHF upgrades onto future facilities and demonstrators}
The leap from HALHF to a multi-TeV linear collider is significant. For example, to achieve \SI{3}{TeV} CoM collisions (the maximum energy envisaged for CLIC) while maintaining a boost of 2.13 would entail \SI{6}{TeV} electrons colliding with \SI{375}{GeV} positrons. Recently the P5~\cite{P52024} report identified \SI{10}{TeV} parton-center-of-mass collisions as the next major milestone for high-energy physics beyond the construction of an \ee\ Higgs factory. If positron acceleration in plasma remains an intractable problem~\cite{Cao2024}, then a high-energy $\gamma\gamma$~\cite{BarklowGessner2023} or even an $\rm{e}^-\rm{e}^-$~\cite{Yakimenko2019} collider 
should be considered. At very high energies, the distinction between \ee, $\gamma\gamma$, $\rm{e}^-\rm{e}^-$, and $\rm{\mu}^+ \rm{\mu}^-$ collisions is blurred due to the dominance of Vector Boson Fusion~\cite{Ali2021}. Therefore, a \SI{10}{TeV} $\gamma\gamma$ collider~\cite{HALHF_upgrades} is considered to be the natural upgrade path for HALHF.

There are several R\&D challenges for a \SI{10}{TeV} linear collider, in addition to the staging that HALHF will have already addressed. Additional challenges include:
\begin{itemize}
\item Extreme beamstrahlung, even with $\gamma\gamma$ collisions;
\item Compton and $\gamma\gamma$ IPs;
\item Betatron radiation in the final plasma-accelerator stages;
\item Offset tolerances;
\item Beam-delivery systems for \SI{5}{TeV} electron beams.
\end{itemize}
A test facility with high-energy beams (100+~GeV) will be necessary to investigate these issues. Such a facility, illustrated in Fig.~\ref{fig:8_HEBTF}, would be possible at ILC or HALHF at \SI{250}{GeV}. In addition, a second IP area could aid development of the Compton and $\gamma\gamma$ IPs.

\section{Plasma-linac design I: staging, driver distribution}
The plasma linac accelerating high-energy electrons for HALHF will require multiple
plasma-accelerator stages. This is a highly non-trivial aspect of the design that requires
re-thinking how to build a high-energy linac, including beam optics, diagnostics,
collimation, beam dumps etc. Fortunately, the plasma linac does not dominate
either the length or the cost, which gives some freedom in how it can be designed. Overall, it is clear that there is much “uncharted territory” and that dedicated R\&D will be
necessary.

\subsection{Staging optics and nonlinear plasma lenses}
\begin{figure}[b]
    \centering
    \includegraphics[width=\linewidth]{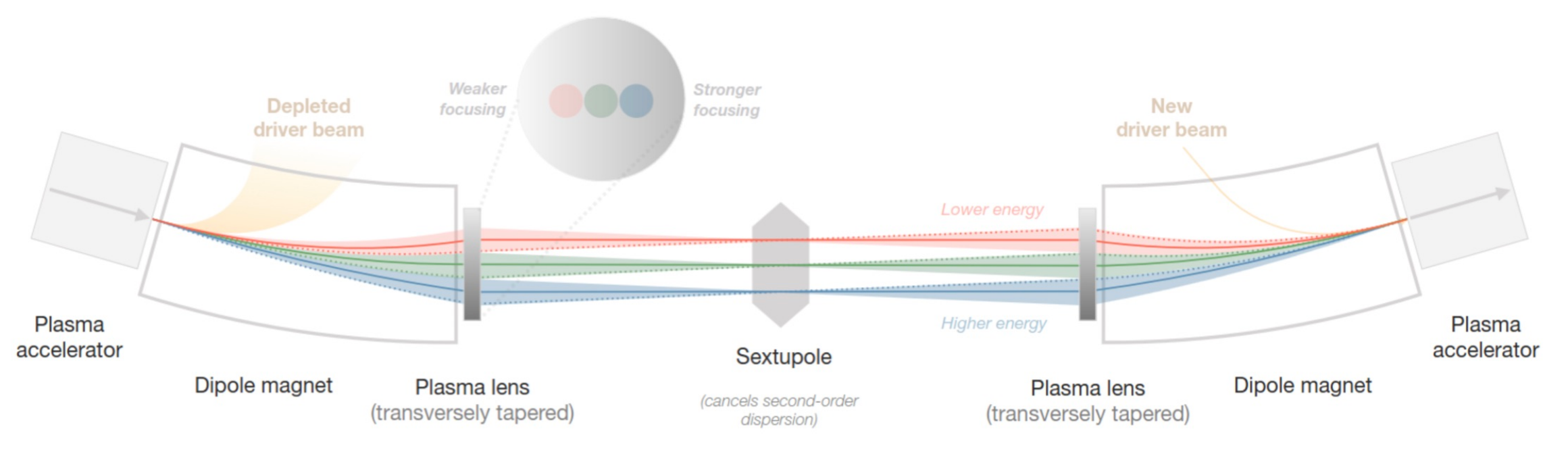}
    \caption{Schematic of an achromatic staging optic based on nonlinear plasma lenses. From Ref.~\cite{Drobniak2024}.}
    \label{fig:9_staging}      
\end{figure}

One of the key requirements of the plasma linac is to deliver stable, low-emittance electron beams, albeit at a less strict level than would be required in symmetric colliders, because of the gain in geometric emittance from working with higher-energy electron beams. While this relaxes emittance preservation in or between every stage, there is nevertheless a strict emittance budget. Chromaticity is the key challenge for the interstage optics, due to the combination of divergence and energy spread in plasma accelerators. The planned solution is to use nonlinear plasma lenses~\cite{Drobniak2024}, which can in principle provide achromatic point-to-point imaging between the plasma-accelerator stages. One question is how the interstage optics transport the non-Gaussian transverse phase-space distribution that arises in the presence of ion motion; the interstage optics should be able to preserve it owing to the achromatic point-to-point imaging (see Fig.~\ref{fig:9_staging}).

The effect of plasma ramps needs to be assessed in this context, as they cause the beam to experience a time-varying ion density due to ion motion. Other important points include emittance growth from wakefield distortions as well as Coulomb scattering in plasma lenses; this must be studied in PIC simulations as well as codes such as \textit{GEANT4}~\cite{Allison2016} to model scattering accurately, as this may result in emittance growth, in particular because the beta function is large inside the plasma lenses. 
Beam–gas scattering inside plasma lenses may well be important and requires a thorough investigation. Finally, the repetition-rate limitations of plasma lenses are not known and may restrict the allowable parameter space of bunch-train patterns in HALHF.

\subsection{Driver distribution}
\label{sec:driver_dist}
A more compact concept for distributing drivers to the plasma stages was discussed, based on a previously presented scheme~\cite{Adli2013}. This involves a periodic, undulating delay chicane with fast kickers (1--\SI{4}{ns} rise time) to extract the rear-most driver into each stage. The proposed setup has two such chicanes: one on each side (left and right) of the plasma accelerator, as this produces an on-average straight multistage linac even when the interstage optics has an angle. In order to bend the drivers enough to delay 1--\SI{2}{ns} per stage, the driver energy is limited to around \SI{10}{GeV}, with smaller values being greatly preferable.  

A topic of discussion was the utility of fast kickers. Given that trains of drivers need to be accelerated in a continuous train without gaps, as this is the most efficient way to produce drivers in an RF linac, the kickers need to have rise and fall times on the ns scale and repeat such kicks hundreds of times with a period that is sub-\SI{100}{ns}. While this is not currently easy to do with strip-line kickers, it may be possible with harmonic RF kickers~\cite{Huang2016}. It was pointed out that having many such complex devices (one per stage) could be troublesome. The stability of these kickers will also need to be very high, as this directly influences the orbit of the driver in the plasma stage. Overall, while it may be possible to use an undulating chicane with fast kickers, another solution is potentially required.

One possible alternative solution separates the drivers into parallel beam lines in a single multi-beam ``splitter” just prior to the plasma stages, requiring more transport lines and delay chicanes, but no fast kickers. It is unclear whether the delay chicanes should bend the drivers in the horizontal or vertical plane. Arguments for bending in the horizontal plane include that this improves the orbit stability in the vertical plane, important for the plasma stages since the vertical emittance has much stricter tolerances. On the other hand, the driver should have a lower emittance in the horizontal plane to avoid resonant emittance mixing~\cite{Diederichs2024}, which is easier to maintain when bending in the vertical. In addition, vertical bending would require less horizontal space, which is always at a premium, in the beam tunnel.

\subsection{Diagnostics between stages}
In order to preserve the emittance between stages and inject beams into the succeeding stage, significant diagnostics must be included between stages. These include: an orbit measurement, using BPMs; an energy-spectrum measurement, likely using an insertable screen at the centre of the interstage optics (close to the central sextupole); and an emittance measurement, possibly using the same screen setup. It was noted that cavity BPMs can in principle measure angles as well as offsets, which could reduce the number of BPMs required. In the currently simulated beam optics, about 10\% of the space is left open for such diagnostics. 

The need for an online/non-destructive measurement of the energy spectrum and emittance was also discussed. This could potentially be achieved using a scintillator screen placed close to the beam, measuring its electric-field halo. Another possibility would be to use the betatron radiation exiting each stage for diagnostic purposes, as the spectrum and spatial distribution of this radiation depend on beam parameters, such as orbit and
emittance.

One possibility that may ease commissioning is to have a dedicated diagnostics and tune-up station between some or all stages. Sending the beam to this station would require flipping the polarity of one of the interstage dipoles. The drivers also require diagnostics as they enter and exit the plasma stages. BPMs should be located before and after each stage, and a natural location for a post-plasma spectrum measurement is the entrance of the driver-beam dump, onto which the energetically dispersed driver is anyway steered.

\subsection{Collimation}
\label{sec:collimation}
While collimation is not a requirement within the plasma linac, in practice multiple plasma accelerators with complex beam optics between them will present a narrower aperture than collimators in the beam-delivery system. This presents a new possibility: to do away with at least some of the (very lengthy) ``conventional” collimator system integrated into the BDS in favour of a distributed collimation system throughout the plasma linac. There are several benefits to such a setup: the  halo furthest distant from the beam core is removed first, at low energy, followed by a gradual removal of the remaining halo at higher energy. 

Two types of collimation will be required in the interstage: energy collimation and beta (transverse-phase-space) collimation. The energy collimation can be performed at the location of the central sextupole, as this location has large dispersion and small beam sizes—ideal for energy collimation. The beta collimation is likely to be done best at the location of the plasma lenses; these anyway impose a radius restriction on the beam. One remaining issue is that of collimating the transverse phase space at different phase advances: the plasma lenses are mainly located at the same phase advance (modulo \SI{180}{\degree}), which makes it challenging to collimate the phase \SI{90}{\degree} away.

\subsection{Beam dumps, heat handling and the radiation environment}
The above discussion on collimation triggered a related discussion: how to deal with the collimated particles. Unless these are directed into dedicated beam dumps, they will lead to heating of the surrounding area, but more importantly activate the plasma-accelerator components. 
The largest issue in this regard is however the dumping of the depleted drivers. Multiple megawatts of dumped beam power---most of it at sub-GeV energies—will need to be handled. A large amount of cooling will also be required. Dedicated radiation simulations using \textit{FLUKA}~\cite{Ballarini2024} are planned to design the appropriate beam dumps.

\section{Plasma-linac design II: polarization, beam-quality preservation, tolerances}
\subsection{Beam-quality preservation and tolerances}
 
Several sources of emittance growth need to be considered for a PWFA collider. Some are well known from RF-based linear-collider studies, including chromatic emittance growth and the wakefield effect due to beam misalignment. In the plasma bubble, the transverse fields, both focusing and the wakefield effects, are stronger than in an RF-collider. The effect of these on the beam therefore needs special attention. For example, transverse wakefields are expected to be very strong, and betatron radiation is an effect that will be important in PWFA-colliders, especially towards TeV-scale energies.  There are also additional effects on the beam in PWFA colliders related to ion motion and the plasma interstages.
 
Tools are being developed to simulate all the relevant effects. The discussion in the workshop started by listing the various known sources of emittance growth in PWFA, and assessing whether the physics of these were captured in the simulation tools used for HALHF.  For the PIC-code \textit{HiPACE++}~\cite{Diederichs2022}, it was assessed that relevant physics for HALHF simulations is currently included, for simulations of a single plasma stage. One issue that needs to be studied further is the need for symmetrical drive beams, which need to be guided to avoid transverse drifts due to misalignment. Multi-stage effects need to be studied along the whole linac, consisting of many stages and interstages. These studies are planned to be done with simplified plasma-stage models in the \textit{ABEL} start-to-end simulation framework, under development at the University of Oslo. One of the features of \textit{ABEL}, which is a multi-level framework which also integrates within it \textit{HiPACE++} and other codes, is a fast wakefield model and betatron-radiation damping which has been benchmarked with \textit{HiPACE++}. An ion-motion model is currently under development.  The interstage part of the simulations is done in the conventional tracking code \textit{ELEGANT}~\cite{Borland2000}, with an additional element modeling an idealised nonlinear active plasma lens.
 
Preliminary results from \textit{ABEL} simulation-scans were presented at the workshop. They indicate that the emittance growth due to transverse wakefields can be sufficiently mitigated with the help of the decoherence effect of ion motion. Studies are ongoing to quantify the tolerances related to wakefields, and also the effect of main-beam misalignment due to drive-beam jitter. The drive beam generates its own focusing axis, and even a very small jitter angle may lead to an unacceptable main-beam misalignment. This effect may be mitigated using guiding magnetic fields around the plasma stage, and will be studied using \textit{ABEL}.
 
\subsection{Spin-polarization preservation}
\label{sec:polarization}
 
Although there is in principle no reason for plasma acceleration not to preserve polarization, this has to be demonstrated in simulations and in experiments.  Published results of simulations~\cite{Vieira2011} indicate already that preservation is possible at the required ILC levels ($>80$\%) at the output of the plasma linac), though the level of preservation depends on the emittance of the accelerated beam.
 
The workshop discussed the level of polarization preservation required for each HALHF stage. A starting objective for the simulation work assumes P=85\% at the start of the plasma linac. The permitted depolarization in each stage is then $(1-0.8/0.85)^{1/n}$, where $n$ is the number of stages, giving e.g.~1.2 permille relative decrease per stage over e.g.~48 stages.  In addition, interstage transport may contribute significantly through synchrotron radiation and nonlinear fields. This needs to be studied and made part of the polarization budget.
 
In an accelerator, the spin will precess according to the Thomas-BMT equations~\cite{Thomas1927, Bargmann1959}, which describe the precession of the spin of a charged particle in electric and magnetic fields. In a plasma bubble (i.e.~the blow-out regime, as used in HALHF) the equations may be greatly simplified, and the spin precession will depend only on the radial focusing force, i.e.~the off-axis position within the bubble. For flat beams ($\sigma_y \ll \sigma_x$), having polarization along the $y$-axis in the plasma seems ideal, as the precession of the $y$-component of spin is minimised. The BDS would then need to rotate the spin to give longitudinal polarization in the collisions.
 
Spin-transport (the simplified Thomas-BMT equations) will be implemented as simplified models in \textit{ABEL}. This will allow for start-to-end simulations, and the level of spin preservation in HALHF may be quantified. Spin-transport is already implemented in \textit{HiPACE++}.
 
While study with simulations is important, facilities/experiments to demonstrate experimentally the conservation of polarization are essential. A challenge is the limited number of polarized beams available for experiments world wide. In the short term, a relatively feasible experiment may be to test polarization preservation in an active plasma lens at ELSA or at MAMi. In the longer term, work on highly polarized plasma photocathodes should continue. An implementation of the SPARTA project with a polarized front-end would be the ultimate demonstrator for HALHF.
 
\section{Plasma generation, heating, cooling and power flow, efficiency}
Plasma accelerators can in principle achieve energy-transfer efficiencies comparable to traditional radio-frequency machines~\cite{Tzoufras2008}. However, due to the nature of the plasma-acceleration process, a certain proportion of the driving beam’s energy will remain in the plasma after the wakefield has passed. What happens to that energy, i.e.~how it is transported and on what timescale, will have implications for the choice of plasma-source technology and how that plasma source is operated. The following subsections examine these topics and their inter-relatedness in the context of HALHF.

\subsection{Efficiency}
In the original HALHF design, the efficiency, i.e.~the energy gained by the accelerating bunch divided by the input driver energy, was 38\%, with 34\% of the initial drive-beam energy remaining in the plasma. Separate experiments have already shown that 57\% of the drive-beam energy can be transferred to the plasma~\cite{Pena2024} and that 42\% of the deposited drive-beam energy can be converted to gain in witness-beam energy~\cite{Lindstrom2021}. Combining these figures suggests an overall efficiency of 24\% is possible, so the HALHF proposal is ambitious but realistic. While a high efficiency of driver energy to the witness beam reduces the heating rate, moderate changes in its value will not meaningfully alter the scale of the technological challenge, which requires plasma devices capable of withstanding orders-of-magnitude higher heating loads than the current state of the art. Fortunately there is the potential to make rapid progress in this area with simple ideas.

\subsection{Plasma generation}
Two common methods of field ionisation were considered: laser or high-voltage-discharge ionisation\footnote{ AWAKE is investigating plasma generation with both discharge and Helicon sources in the range 1--\SI{10e14}{\per\cubic\cm} with a baseline of \SI{7e14}{\per\cubic\cm}~\cite{Buttenschon2018, Ariniello2019, Torrado2023}}. The target density now envisaged for HALHF is \SI{\sim e15}{\per\cubic\cm}, which is compatible with both generation mechanisms. A pre-ionised plasma source, rather than relying on beam ionisation, is preferred to combat driver head erosion. Laser ionisation is potentially the more flexible of the two options, as advanced shaping methods (with the use of Bessel beams for example) may give control over the shape of the plasma density ramps, which are crucial for emittance preservation~\cite{Chen2015}. Additionally, plasma of any (sub-critical) density can be ionised. In order to ionise Ar or H, a focused intensity of around \SI{2e14}{W cm^{-2}} is required. The laser energy of a \SI{40}{fs} (Ti:sapphire) laser required to ionise a \SI{5}{m} long stage was calculated assuming a plasma column radius of twice the blowout radius (assumed to be \SI{0.42}{mm}). A Bessel beam to do this would require 8 rings, each of energy \SI{55}{mJ}. Therefore, a laser delivering \SI{>0.5}{J} (plus some conservative safety factor) would be required, which at \SI{10}{kHz} is far beyond the state of the art. Such lasers also typically run with constant (CW) pulse spacing, perhaps excluding compatibility with pulsed operation. However, in the next section, it will be shown that the energy deposition from the train of driver bunches should keep the plasma hot enough to remain fully ionised, meaning that plasma confinement may be a more pressing issue than ultrahigh-repetition-rate plasma generation, and that only one external ionisation event may be required per bunch train. 

Discharge ionisation is relatively simple, although the downside is that it favours higher densities (to stay close to the Paschen curve minimum in a 5 m plasma, a density of order \SI{e16}{\per\cubic\cm} is required). Mitigation strategies exist such as using a hot gas, employing a glow discharge or localised laser ionisation to initiate the discharge. Scaled calculations from FLASHForward measurements~\cite{Loisch2025} indicate that the plasma required for HALHF can be produced with a few tens of Joules. The ease and low energy requirements of discharge ionisation recommend it for the HALHF baseline design.

\subsection{Heating}
The original HALHF parameters anticipate a beyond-state-of-the-art yet realistic overall energy-transfer efficiency. However, even at this working point, 34\% of the drive-beam energy will be deposited in the plasma. This equates to \SI{\sim 50}{J} of total energy per acceleration event (or \SI{\sim 10}{J/m} averaged over the length of the plasma-acceleration module) reaching \SI{50}{kJ} from the full 100-bunch train. Unfortunately, relatively little is known about how this energy is transported within the plasma but naïve assumptions can help frame the problem. For example, if all the energy deposited in the plasma is equally distributed to all plasma electrons and ions in the plasma source, the temperature of the plasma constituents will rise to \SI{\sim 2}{keV}, assuming a small capillary radius of order mm, from a single bunch—sufficiently high to ionise almost all levels of argon. Furthermore, if no energy were to be lost from the plasma between bunches in the train via, for example, electromagnetic radiation or conduction by the plasma source, this temperature could rise to \SI{\sim 200}{keV}, an order of magnitude hotter than some fusion reactors. Lower temperatures may be reached by using cm-scale plasmas, if the energy can be evenly distributed on a sufficiently short timescale. Ultimately, little is known about the operation of plasma accelerators at these temperatures, thus targeted research, in the form of long-term PIC simulations, with all the necessary physics included, and experimental results from plasma-electron and -ion temperature diagnostics, are urgently required. 

\subsection{Cooling} 
The energy deposited in the plasma will make its way to the surrounding plasma vessel if no remedial measures are taken. Although the energy-transport channels in a plasma accelerator remain relatively unmapped, upper bounds on the required cooling rates can be calculated from the energy deposited in the plasma from the drive beam. The HALHF baseline parameters need around $10^4$ bunches per second to achieve the necessary luminosity. With the \SI{10}{J/m} of energy deposition previously calculated for a single acceleration event, the time-averaged cooling rate required for the entire bunch train corresponds to 100 kW/m, an order of magnitude higher than the cooling rates expected at CLIC. Furthermore, if HALHF were to operate in a burst mode as originally proposed, the energy deposition over the bunch train would be \SI{50}{kJ/m} in \SI{10}{\micro\s}, likely leading to MPa principle stresses on the material housing the plasma. Novel plasma-source designs capable of withstanding the extreme and rapid stresses and temperature changes are therefore an important R\&D topic for HALHF.

\begin{figure}[!ht]
    \centering
    \includegraphics[width=\linewidth]{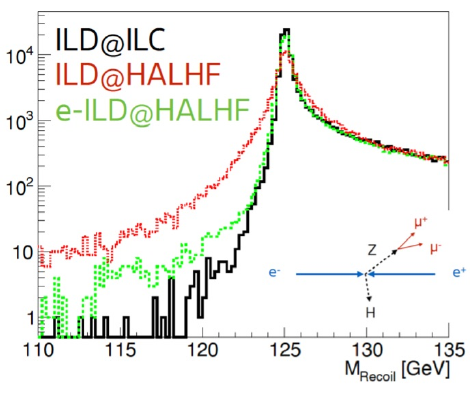}
    \caption{Recoil mass distribution in the process \ee $\rightarrow$ ZH $\rightarrow \rm{\mu\mu} \rm{H}$ events in the ILD detector at HALHF (red line) compared to the same detector at ILC (black line). Elongation of the barrel by a factor 2 is shown in by the green line.}
    \label{fig:10_ZH}       
\end{figure}

\begin{table*}
\vskip-130pt
\hskip-50pt
  \scriptsize
    \begin{tabular}{p{0.28\linewidth}>{\centering}p{0.06\linewidth}>{\centering}p{0.08\linewidth}>{\centering\arraybackslash}p{0.08\linewidth}>{\centering\arraybackslash}p{0.08\linewidth}>{\centering\arraybackslash}p{0.08\linewidth}>{\centering\arraybackslash}>{\centering\arraybackslash}p{0.08\linewidth}>{\centering\arraybackslash}p{0.08\linewidth}}
    
    \textit{Machine parameters} & \textit{Unit} & 
   \multicolumn{2}{c}{\textit{Value (250 GeV)}}& \multicolumn{2}{c}{\textit{Value (380 GeV)}} & \multicolumn{2}{c}{\textit{Value (550 GeV)}}\\
    \hline
    Center-of-mass energy & GeV & \multicolumn{2}{c}{250} & \multicolumn{2}{c}{380} &\multicolumn{2}{c}{550}\\
    Center-of-mass boost & ~ & \multicolumn{2}{c}{1.67} & \multicolumn{2}{c}{1.67} & \multicolumn{2}{c}{1.67}\\
    Bunches per train & & \multicolumn{2}{c}{160} & \multicolumn{2}{c}{160}  & \multicolumn{2}{c}{160}  \\
    Train repetition rate & Hz & \multicolumn{2}{c}{100} & \multicolumn{2}{c}{100} & \multicolumn{2}{c}{100}\\
    Average collision rate & kHz & \multicolumn{2}{c}{16} & \multicolumn{2}{c}{16} & \multicolumn{2}{c}{16}\\
    Luminosity & cm$^{-2}$~s$^{-1}$ & \multicolumn{2}{c}{$1.2\times10^{34}$} & \multicolumn{2}{c}{$1.7\times10^{34}$} & \multicolumn{2}{c}{$2.5\times10^{34}$} \\
    Luminosity fraction in top 1\% & & \multicolumn{2}{c}{63\%} & \multicolumn{2}{c}{53\%} &\multicolumn{2}{c}{46\%} \\
    Quantum parameter ($\Upsilon$) & & \multicolumn{2}{c}{0.9} & \multicolumn{2}{c}{1.6} &\multicolumn{2}{c}{2.8} \\
    Estimated total power usage & MW & \multicolumn{2}{c}{106} & \multicolumn{2}{c}{154} & \multicolumn{2}{c}{218} \\
        Total site length & km & \multicolumn{2}{c}{4.9} & \multicolumn{2}{c}{6.5} & \multicolumn{2}{c}{8.4} \\
    \hline
     & & & \\[-8pt]
    \textit{Colliding-beam parameters} & & $e^{-}$ & $e^{+}$ & $e^{-}$ & $e^{+}$ & $e^{-}$ & $e^{+}$ \\ 
    \hline
    Beam energy & GeV & 375 & 41.7 & 570 & 63.3 & 825 & 91.7 \\
    Bunch population & $10^{10}$ & 1 & 3 & 1 & 3 & 1 & 3 \\
    Bunch length in linacs (rms) & $\mu$m & 40 & 150 & 40 & 150 & 40 & 150 \\ 
    Bunch length at IP (rms) & $\mu$m & \multicolumn{2}{c}{150} & \multicolumn{2}{c}{150} & \multicolumn{2}{c}{150}  \\
    Energy spread (rms) & \% & \multicolumn{2}{c}{0.15} & \multicolumn{2}{c}{0.15} & \multicolumn{2}{c}{0.15} \\
    Horizontal emittance (norm.) & $\mu$m & 90 & 10 & 90 & 10 & 90 & 10 \\
    Vertical emittance (norm.) & $\mu$m & 0.32 & 0.035 & 0.32 & 0.035 & 0.32 & 0.035\\
    IP horizontal beta function & mm & \multicolumn{2}{c}{3.3} & \multicolumn{2}{c}{3.3} & \multicolumn{2}{c}{3.3}\\
    IP vertical beta function & mm & \multicolumn{2}{c}{0.1} & \multicolumn{2}{c}{0.1} & \multicolumn{2}{c}{0.1} \\
    IP horizontal beam size (rms) & nm & \multicolumn{2}{c}{636} & \multicolumn{2}{c}{519} & \multicolumn{2}{c}{429} \\
    IP vertical beam size (rms) & nm & \multicolumn{2}{c}{6.6} & \multicolumn{2}{c}{5.2} & \multicolumn{2}{c}{4.4} \\
    Average beam power delivered & MW & 9.6 & 3.2 & 14.6 & 4.9 & 21.1 & 7.0 \\
    Bunch separation & ns & \multicolumn{2}{c}{16} &\multicolumn{2}{c}{16} &\multicolumn{2}{c}{16} \\
    Average beam current & {\textmu}A & 26 & 77 & 26 & 77 & 26 & 77  \\[2pt]
    \hline \\[-6pt]
    \multicolumn{2}{l}{\textit{Positron cool-copper RF linac parameters (S-band)}}  & & & & \\
     \hline
     Average cavity gradient & MV/m & \multicolumn{2}{c}{40}& \multicolumn{2}{c}{40} & \multicolumn{2}{c}{40}\\
        Average gradient & MV/m & \multicolumn{2}{c}{36} & \multicolumn{2}{c}{36} & \multicolumn{2}{c}{36}\\
    Wall-plug-to-beam efficiency & \% & \multicolumn{2}{c}{11} & \multicolumn{2}{c}{11} & \multicolumn{2}{c}{11}\\
    RF power & MW & \multicolumn{2}{c}{11.7} &\multicolumn{2}{c}{17.8} &\multicolumn{2}{c}{25.8} \\
    Cooling power  & MW & \multicolumn{2}{c}{17.9} &  \multicolumn{2}{c}{27.3} & \multicolumn{2}{c}{39.5}\\
     Total power & MW & \multicolumn{2}{c}{29.6} & \multicolumn{2}{c}{45.1} & \multicolumn{2}{c}{65.3} \\
   Klystron peak power & MW & \multicolumn{2}{c}{67} & \multicolumn{2}{c}{67} & \multicolumn{2}{c}{67} \\
     Number of klystrons & ~ & \multicolumn{2}{c}{321} & \multicolumn{2}{c}{452} & \multicolumn{2}{c}{678} \\
    RF frequency & GHz & \multicolumn{2}{c}{3} & \multicolumn{2}{c}{3} & \multicolumn{2}{c}{3} \\
 Operating Temperature & K & \multicolumn{2}{c}{77} & \multicolumn{2}{c}{77} & \multicolumn{2}{c}{77} \\
         Length (after damping ring, starting at 3 GeV) & km & \multicolumn{2}{c}{1.1} & \multicolumn{2}{c}{1.7} & \multicolumn{2}{c}{2.5} \\
    \hline \\[-6pt]
    \multicolumn{2}{l}{\textit{Driver linac RF parameters (L-band)}} & & & & \\
    \hline \\[-6pt]
        Average cavity gradient & MV/m & \multicolumn{2}{c}{4} & \multicolumn{2}{c}{4} \\
    Average gradient & MV/m & \multicolumn{2}{c}{3} & \multicolumn{2}{c}{3} & \multicolumn{2}{c}{3} \\
    Wall-plug-to-beam efficiency & \% & \multicolumn{2}{c}{55} & \multicolumn{2}{c}{55} & \multicolumn{2}{c}{55} \\
    RF power usage & MW & \multicolumn{2}{c}{42.9} & \multicolumn{2}{c}{66.0} & \multicolumn{2}{c}{96.4}\\
   Klystron peak power & MW & \multicolumn{2}{c}{21} &  \multicolumn{2}{c}{21} & \multicolumn{2}{c}{21}\\
     Number of klystrons & ~ & \multicolumn{2}{c}{409} & \multicolumn{2}{c}{630} & \multicolumn{2}{c}{919} \\
    RF frequency & GHz & \multicolumn{2}{c}{1} & \multicolumn{2}{c}{1} & \multicolumn{2}{c}{1} \\
        Length & km & \multicolumn{2}{c}{1.3} & \multicolumn{2}{c}{1.9} & \multicolumn{2}{c}{2.8} \\
\hline \\[-6pt]
\multicolumn{2}{l}{\textit{Combiner Ring parameters}} 
    & & \\
    \hline \\[-6pt]
    ~ & ~ & ~ & ~ & ~ & ~ \\[-8pt]
    Delay loop length & m & \multicolumn{2}{c}{1.5} &\multicolumn{2}{c}{1.5} &\multicolumn{2}{c}{1.5} \\
    CR1 diameter & m & \multicolumn{2}{c}{244} &\multicolumn{2}{c}{244} &\multicolumn{2}{c}{244} \\
    CR2 diameter & m & \multicolumn{2}{c}{244} &\multicolumn{2}{c}{244} &\multicolumn{2}{c}{244} \\
    \hline \\[-6pt]
\multicolumn{2}{l}{\textit{PWFA linac and drive-beam parameters}} 
    & & & & \\
    \hline \\[-6pt]
    Number of stages &  & \multicolumn{2}{c}{48} & \multicolumn{2}{c}{48} & \multicolumn{2}{c}{48} \\
    Plasma density & cm$^{-3}$ & \multicolumn{2}{c}{$6\times10^{14}$} & \multicolumn{2}{c}{$6\times10^{14}$} & \multicolumn{2}{c}{$6\times10^{14}$}\\ 
    In-plasma accel. gradient & GV/m & \multicolumn{2}{c}{1}& \multicolumn{2}{c}{1} & \multicolumn{2}{c}{1} \\
    Av. gradient (incl. optics) & GV/m & \multicolumn{2}{c}{0.33}& \multicolumn{2}{c}{0.38} & \multicolumn{2}{c}{0.43} \\
    Transformer ratio & ~ & \multicolumn{2}{c}{2} & \multicolumn{2}{c}{2} & \multicolumn{2}{c}{2} \\
    Length per stage & m & \multicolumn{2}{c}{7.8} &\multicolumn{2}{c}{11.8} & \multicolumn{2}{c}{17.1} \\
    Energy gain per stage\tnote{a} & GeV & \multicolumn{2}{c}{7.8} & \multicolumn{2}{c}{11.8} & \multicolumn{2}{c}{17.1} \\
    Initial injection energy & GeV & \multicolumn{2}{c}{3} & \multicolumn{2}{c}{3} & \multicolumn{2}{c}{3}\\
    Driver energy & GeV & \multicolumn{2}{c}{4}  & \multicolumn{2}{c}{5.9} & \multicolumn{2}{c}{8.6}\\
    Driver bunch population & $10^{10}$ & \multicolumn{2}{c}{5.0} & \multicolumn{2}{c}{5.0} & \multicolumn{2}{c}{5.0} \\
    Driver bunch length (rms)  & $\mu$m &  \multicolumn{2}{c}{253} &  \multicolumn{2}{c}{253} & \multicolumn{2}{c}{253} \\
    Driver average beam power & MW & \multicolumn{2}{c}{23.8} & \multicolumn{2}{c}{36.2} & \multicolumn{2}{c}{52.6} \\
    Driver bunch separation & ns & \multicolumn{2}{c}{4} & \multicolumn{2}{c}{4} & \multicolumn{2}{c}{4} \\
    Driver-to-wake efficiency & \% & \multicolumn{2}{c}{80} & \multicolumn{2}{c}{80} & \multicolumn{2}{c}{80} \\
    Wake-to-beam efficiency & \% & \multicolumn{2}{c}{50} & \multicolumn{2}{c}{50} & \multicolumn{2}{c}{50} \\
    Driver-to-beam efficiency & \% & \multicolumn{2}{c}{40} & \multicolumn{2}{c}{40} & \multicolumn{2}{c}{40} \\
    Wallplug-to-beam efficiency & \% & \multicolumn{2}{c}{22} & \multicolumn{2}{c}{22} & \multicolumn{2}{c}{22} \\
    Cooling req.~per stage length & kW/m & \multicolumn{2}{c}{38.4} & \multicolumn{2}{c}{38.4} & \multicolumn{2}{c}{38.4} \\
    Length & km & \multicolumn{2}{c}{1.1} & \multicolumn{2}{c}{1.5} & \multicolumn{2}{c}{1.9} \\
    \hline
    \end{tabular}
   \caption{HALHF parameters for the updated baseline design at \SI{250}{GeV}, \SI{380}{GeV} and and \SI{550}{GeV} CoM energies.}
   \label{tab:2}
\end{table*}

\section{Physics/detector design \& constraints, including coherent pairs}
The detector design for symmetric Higgs factories has been optimised with a focus on the central detector region. In the forward and backward parts of the detector, typical performance criteria like the momentum resolution, the vertex resolution and the jet energy resolution become worse. At HALHF, all collision events will be boosted into the forward direction, i.e.~the direction of the electron beam. Thus, placing a typical symmetric Higgs factory detector at the HALHF interaction point would lead to significantly worse precision on flagship measurements. Figure~\ref{fig:10_ZH} shows a clear and unacceptable deterioration of the recoil mass distribution in \ee $\rightarrow$ ZH $\rightarrow \mu\mu \rm{H}$ events in the ILD detector placed at HALHF compared to the same detector at ILC. The loss in performance is primarily due to a shortened lever-arm in the solenoidal magnetic field of the detector, and can be recovered e.g.~by lengthening the barrel part of the detector by a factor two. This is not intended to be a serious proposal for HALHF, merely to indicate that a properly designed detector can recover similar resolution to that at a symmetric Higgs factory.

The vertex reconstruction suffers from reduced vertex detector coverage at small angles in the ILD. However, the boost at HALHF is expected to facilitate vertex reconstruction since the flight distances in the lab system become longer for the same proper lifetime of a particle. Quantifying these effects requires an optimised geometry of the beam pipe, the vertex detector and the forward tracking. This is subject to one further constraint, namely the distribution of \ee\ pairs created in the beam-beam interaction. The amount and distribution of the pair background has been studied as a function of the collision parameters. It turns out that the size of the region which must be free of instrumentation is driven by the charge asymmetry of the beams, rather than the energy asymmetry. Furthermore, the bunch lengths and the strength of the detector solenoid, which acts to sweep away the \ee\ pairs, also influence the region that can be instrumented. The new baseline will be used to optimise the inner and forward parts of a purpose-designed detector for HALHF and the result will be implemented in a detailed, \textit{GEANT4}-based detector simulation for realistic studies of the vertex-reconstruction performance. 

The luminosity measurement at \ee\ colliders classically relies on low-angle Bhabha scattering, i.e.~back-to-back \ee\ pairs at small polar angles, detected in special high-resolution luminosity calorimeters. Due to the steep fall of the rate with the polar angle, not only the energy but also the position measurement of the clusters is crucial, thus limiting factors are typically the alignment of the two luminosity calorimeters as well as deflection of the outgoing $\rm{e^+} / \rm{e^-}$ in the strong electromagnetic field of the colliding bunches. At HALHF, the Bhabhas will not be back-to-back, but the $\rm{e^-}$ will be boosted even more forward while the $\rm{e^+}$ ends up at a fixed polar angle in the barrel region of the detector. While no fundamental show-stopper arises from this configuration in principle, a detailed simulation study has to be conducted in order to verify the performance and evaluate requirements on the detector.

\section{New baseline}

One upgrade with respect to the original baseline is the provision for positron polarisation, which was previously only an option. The physics gain implied by polarisation of both leptons is sufficiently significant to outweigh the complications inherent in producing polarised positrons. The design is based on that of ILC, in which the colliding electron beam at full energy is passed through a wiggler and the resulting photons impinge on a rotating target. The positrons from the resulting \ee\ pairs are concentrated and collected. With appropriate modifications to the wiggler parameters, it is believed that a high degree of polarization can be achieved.

The decision taken at this workshop to separate out the drive-beam and positron linacs gives a much wider scope for optimisation of the overall HALHF facility. In particular, the removal of the necessity to have the same energy for the drive-beams and the colliding positron beam is important. The enlarged parameter space was explored and optimised using a multidimensional Bayesian optimisation programme. Input to this was a detailed cost-model, some parts of which remain preliminary but these are not thought  to affect the optimisation process significantly. The metric used in the optimisation consisted of the sum of separate costs for construction, overheads, running costs, maintenance over the duration of the full programme and a ``carbon tax" to give weight to minimising the carbon footprint of the facility. The full programme cost was optimised to produce a total integrated luminosity of \SI{2}{ab^{-1}} at \SI{250}{GeV} CoM energy. It was also possible to minimise the facility construction cost without regard to running costs or duration of such a programme. In practice these two minima were relatively close together and quite shallow in parameter space. The final parameter set varied from either by using ``by hand" adjustments that were intended to make the facility more optimal using a variety of less tangible criteria unsuitable for quantification in an optimiser (increasing the overall cost by less than 5\%). 

\begin{figure}[!h]
    \centering
    \includegraphics[width=\linewidth]{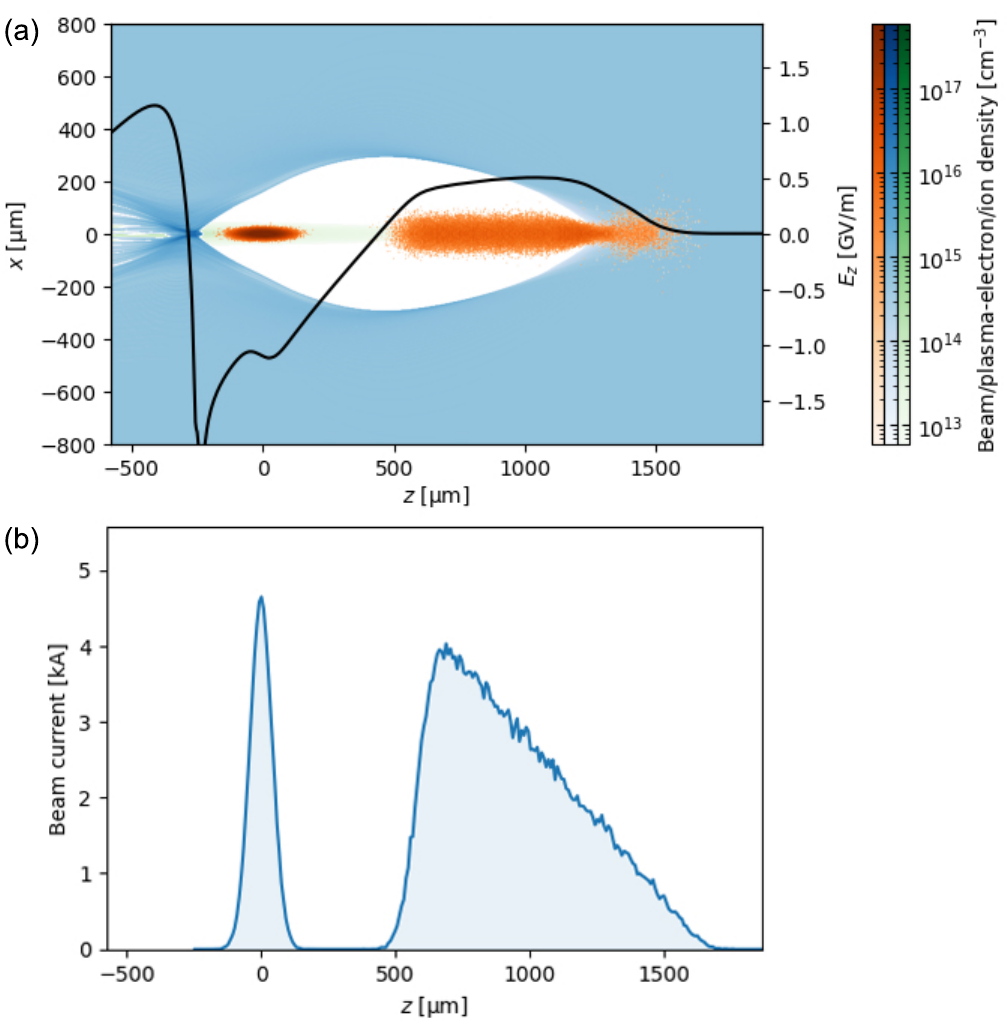}
    \caption{(a) First time step of a PIC simulation with new baseline parameters, performed using HiPACE++. The plasma-electron density (blue colormap) is \SI{6e14}{\per\cm\cubed}, and the driver and main beam (orange colormap) has a charge of 8 and 1.6 nC, respectively. A helium plasma is used in this simulation, resulting in a controlled amount of excess ion density on axis (green colormap). (b) The corresponding current profile shows a peak current around 4--\SI{5}{kA}, with a ramped driver current and an initially Gaussian main beam (prior to self correction).}
    \label{fig:11_PIC_simulation}
\end{figure}

\begin{figure*}[!h]
\centering\includegraphics[angle=90, scale=0.45]{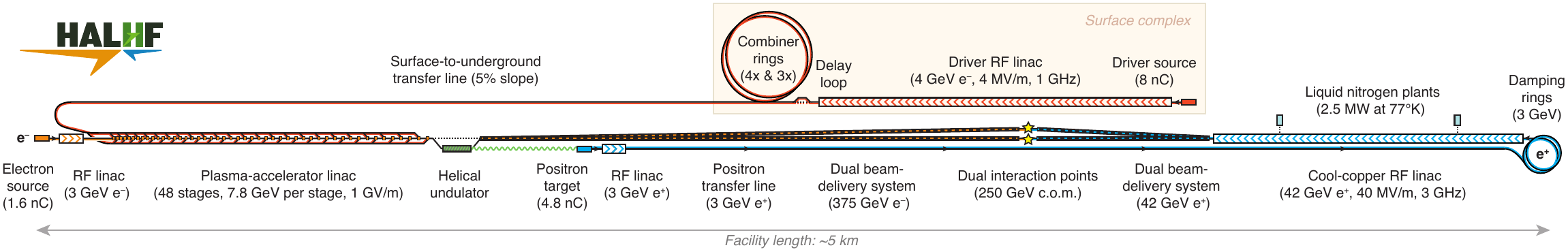}
    \caption{Schematic view of the new HALHF baseline, using a cool-copper positron linac. The red sections relate to electrons, blue to positrons and green to photons.}
    \label{fig:12_HALHF_cold}       
\end{figure*}

Balancing the power requirement (and thereby the number of klystrons plus modulators, which are the largest cost element) between the drive-beam linac and that accelerating the colliding positron bunch is one of the key parameter spaces explored in the Bayesian optimisation. Another is the length, directly proportional to the assumed accelerating gradient, of the linac structures. It transpires that the gradient of the PWFA arm is not a strong cost driver, provided that the gradient in each plasma cell remains at least approximately an order of magnitude larger than the conventional linacs. Thus the reduction of the gradient from that of the original baseline via a reduction in the plasma density by an order of magnitude to \SI{6e14}{\per\cm\cubed} is selected (see Fig.~\ref{fig:11_PIC_simulation}), which greatly reduces performance requirements on many aspects of the PWFA accelerator, which is the least understood element of the facility. This is a very conservative choice; once experience is gained with operating a PWFA accelerator in a collider, it should be possible to increase the gradient, and thereby reduce the length (or increasing the electron energy), by increasing the plasma density. The optimum point found by the Bayesian optimiser using the new layout and parameters reduces the energy asymmetry significantly from that of the original baseline. The new baseline has an electron energy of 375~GeV and a positron energy of \SI{41.7}{GeV}, corresponding to a boost of 1.67. The original baseline had a boost of 2.13. A schematic of the new baseline, which also includes provision for two IRs, is shown in Fig.~\ref{fig:12_HALHF_cold}.

The optimisation of the separate drive-beam linac quickly converges to a system very close to that designed for the CLIC drive beam~\cite{CLIC_CDR_2012}. This is reassuring, as the latter has been the subject of many person-years of design and the requirements are similar to those of HALHF. These linacs can be thought of as one arm of a transformer that converts low-energy, high-current electron bunches to high-energy and low-current bunches. The charge in each drive bunch in the HALHF linac is \SI{8}{nC} and the energy \SI{4}{GeV}. The similarity between the new HALHF baseline and CLIC extends to the addition in HALHF of a delay loop plus combiner ring, with exactly the same functions, \textit{viz.} to produce the required bunch pattern for HALHF while reducing the peak power load on the linac to an acceptable level. The drive-beam linac has an average gradient of \SI{\sim 3}{MV/m}, higher than that of CLIC of \SI{\sim 1}{MV/m}, and runs with \SI{1}{GHz} klystrons and modulators. This leads to somewhat more resistive losses than in CLIC, but the RF-to-driver efficiency still remains above 90\% due to high beam loading. Because of the higher driver energy as well as a longer bunch separation, the combiner rings are larger than those of CLIC. The two combiner rings are assumed to be in the same tunnel and interleave in two steps of 3 and 4, giving a combination factor of 12. Every fourth RF bucket is filled. Other parameters can be seen in Table~\ref{tab:2}.

The PWFA linac has a larger number of stages than the original baseline: 48 compared to 16. Each has correspondingly lower gain. This choice is related to overheads in e.g.~the combiner rings arising from using higher-energy drive beams and because the power efficiency of high-current, low-gradient RF accelerators is higher than vice versa (also a major design driver in CLIC). The length of each stage is \SI{7.8}{m} and the gradient per cell is \SI{1}{GV/m}. The drive beams are distributed to each cell in synchronisation with the accelerated bunch by a system of RF deflectors and fast kickers in which drive bunches emerging from the combiner rings are alternately distributed to each side of the array of plasma cells. The spent drive beams, which have a near-100\% energy spread but spiked at low energy (a few \SI{100}{MeV}), are extracted to beam dumps after each cell. On reaching the desired energy, the beams are directed to the BDS. 

\begin{figure*}[!h]
\centering\includegraphics[angle=90, scale=0.38]{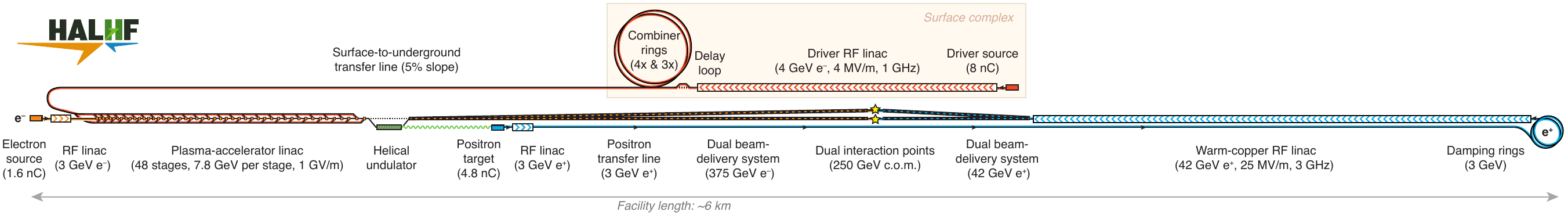}
    \caption{Schematic view of the new HALHF baseline with the fall-back warm positron linac. Other details as in the caption to Fig.~\ref{fig:12_HALHF_cold}.}
    \label{fig:13_HALHF_warm}       
\end{figure*}

The positron linac baseline is S-band, using copper structures cooled to liquid-nitrogen temperatures, as developed by the $\rm{C}^3$ Collaboration~\cite{C3}.  This is chosen because of the high gradient that can be produced, reducing the length of the linac, which otherwise would be the longest single item in the HALHF facility. A very conservative \SI{40}{MV/m} gradient is assumed, much less than that required for the $\rm{C}^3$ Higgs factory. The choice of cooled copper adds complication in that liquid-nitrogen cryogenic is required, with a power of \SI{16.6}{MW}, with concomitant cryogenic plants on the surface and the safety implications of liquid nitrogen in the tunnel. While this technology is novel, its use in HALHF is conservative in comparison to the requirements for a $\rm{C}^3$ Higgs factory. A fallback has nevertheless been designed using a warm linac with parameters very similar to the original and venerable SLAC linac, as shown in Fig.~\ref{fig:13_HALHF_warm}. This is assumed to have a gradient of \SI{25}{MV/m}. 

The beam-delivery systems are scaled based on that developed for the ILC up to a beam energy of \SI{1}{TeV}. As discussed in Sec.~\ref{sec:bds-positron}, various novel schemes for the BDS are under consideration, which will hopefully either reduce its length or allow it to cope with higher energy upgrades for the same length. This is particularly important for the electron arm, where collimation between plasma cells could well play an important role. Twin beam-delivery systems are envisaged, allowing a sharing of luminosity between two detectors. Currently it is envisaged that the detectors would share a single hall but a final decision will be taken when a detailed BDS design has been carried out. 

As remarked earlier, some elements of HALHF have hardly been studied at all due to lack of resources. These include the positron damping rings, where it has been assumed that variants on the CLIC design can be used, and the electron and driver sources. Here it has been assumed that R\&D currently underway can succeed in developing e.g.~robust and long-lived polarized electron sources that can give the required HALHF beam parameters without the necessity for a damping ring~\cite{Maxon2024}. Should this transpire not to be the case, a damping ring can be added as a relatively minor upgrade. 

Although the capital cost and running costs are output as a result of the Bayesian optimisation, it would be premature to release estimates of either at the current stage. The combination of an optimisation of the \textit{total} cost of the project to produce an integrated luminosity of \SI{2}{ab^{-1}}, the addition of a separate positron linac, and upgrades to a polarized-positron source and two interaction points lead to an increase in capital cost. An approximate estimate is that the capital cost of the new baseline will be approximately a factor 1.4--1.8 greater than the original; however, the power budget will be very similar.

\section{Summary \& Outlook}
\label{sec:Conclusions}
The Erice HALHF workshop represented a concentrated deep dive into most aspects of the performance and technology of the HALHF project in beautiful surroundings conducive to the task in hand. As often happens in such situations, perspectives were changed and completely new questions and concerns arose. As stated in the original publication, some of the main benefits of HALHF arise from the working through of questions that are only posed in a concrete implementation of a new technology. In the case of PWFA, many features, such as the vertical--horizontal mixing in flat beams due to ion motion~\cite{Diederichs2024}, only arose because HALHF requires a specific implementation. The discussions in the workshop convinced the collaboration that a separation of positron and drive-beam linacs was desirable, particularly with regard to the applicability of HALHF to TeV energies. Other improvements were identified in the course of the discussions. These, together with the scope increases involved in the provision of two IPs and positron polarization, will lead to an increase in the HALHF capital cost. The use of Bayesian optimisation gives confidence that the choice of parameters for the new HALHF baseline is close to the optimum. 

With this new baseline, the HALHF collaboration intends to refine those aspects of the project for which it has the expertise and effort available in order to produce the strongest and most convincing input to the update to the European Particle Physics Strategy. Many of the aspects, in particular those of the PWFA arm, need substantial further R\&D. Without any substantial resources devoted to HALHF, progress has had to rely substantially on individuals freeing up time and parasitic work on other projects. As a result the minimum ten-year R\&D programme outlined in the original HALHF publication~\cite{HALHF} remains a ten-year programme eighteen months later. Nevertheless, some projects, such as SPARTA~\cite{SPARTA}, are highly relevant to many aspects of HALHF, in particular staging of multiple plasma cells. The collaboration is excited by the progress being made in this project.

The HALHF Collaboration remains enthusiastic and convinced that HALHF is an exciting and novel project that at the least will have substantial spin-offs of plasma-accelerator technology applicable in many fields. In particle physics, the collaboration believes that HALHF would be cheaper and have a smaller carbon footprint than other Higgs-factory implementations and as such will continue to develop it further.

%% ACKNOWLEDGEMENTS
\vspace{0.5cm}
\section*{Acknowledgments}
We are deeply grateful to the Ettore Majorana Foundation and Centre for Scientific Culture for their hospitality and use of their facilities during this workshop. Its success was due in no small part to these excellent facilities and surroundings.

The HALHF Collaboration is grateful for the support of its funding authorities. In particular, this work was supported by the Research Council of Norway (NFR Grant No.~313770), the European Research Council ERC-2023-StG call under grant agreement No 101116161 SPARTA. BF is grateful for the support from the Leverhulme Trust via an Emeritus Fellowship; and he and other colleagues from the John Adams and Cockcroft Institutes and Rutherford Appleton Laboratory acknowledge support from the Science and Technology Facilities Council, UK. This research was funded in whole, or in part, by the UKRI. For the purpose of Open Access, the authors have applied a CC BY public copyright licence to any Author Accepted Manuscript version arising from this submission.The authors are also grateful for continuing support from DESY. Authors from institutions in the United States of America acknowledge the support of the US Department of Energy. 

\bibliographystyle{elsarticle-harv}

\end{document}